\documentclass{iopart}

\usepackage{iopams}
\usepackage{subfig}
\usepackage{graphicx}
\usepackage{color}
\usepackage{hyperref}
\expandafter\let\csname equation*\endcsname\relax
\expandafter\let\csname endequation*\endcsname\relax
\usepackage{amsmath}
\usepackage{lineno}

\begin{document}

\title[]
{The KAGRA underground environment and lessons for the Einstein Telescope}

\author{Francesca Badaracco}
\address{Gran Sasso Science Institute (GSSI), I-67100 L'Aquila, Italy}
\address{INFN, Laboratori Nazionali del Gran Sasso, I-67100 Assergi, Italy}
\author{Camilla De Rossi}
\address{European Gravitational Observatory (EGO), I-56021 Cascina, Pisa, Italy}
\author{Irene Fiori}
\address{European Gravitational Observatory (EGO), I-56021 Cascina, Pisa, Italy}
\author{Jan Harms}
\address{Gran Sasso Science Institute (GSSI), I-67100 L'Aquila, Italy}
\address{INFN, Laboratori Nazionali del Gran Sasso, I-67100 Assergi, Italy}
\author{Kouseki Miyo}
\address{Institute for Cosmic Ray Research (ICRR), KAGRA observatory, The University of Tokyo, Kamioka-cho, Hida Ciry, Gifu 506-1205, Japan}
\author{Federico Paoletti}
\address{INFN, Sezione di Pisa, I-56127 Pisa, Italy}
\author{Taiki Tanaka}
\address{Institute for Cosmic Ray Research (ICRR), KAGRA observatory, The University of Tokyo, Kamioka-cho, Hida Ciry, Gifu 506-1205, Japan}
\author{Tatsuki Washimi}
\address{Gravitational Wave Science Project (GWPS), Kamioka branch, National Astronomical Observatory of Japan (NAOJ), Kamioka-cho, Hida Ciry, Gifu 506-1205, Japan}
\author{Takaaki Yokozawa}
\address{Institute for Cosmic Ray Research (ICRR), KAGRA observatory, The University of Tokyo, Kamioka-cho, Hida Ciry, Gifu 506-1205, Japan}

\begin{abstract}
The KAGRA gravitational-wave detector in Japan is the only operating detector hosted in an underground infrastructure. Underground sites promise a greatly reduced contribution of the environment to detector noise thereby opening the possibility to extend the observation band to frequencies well below 10\,Hz. For this reason, the proposed next-generation infrastructure Einstein Telescope in Europe would be realized underground aiming for an observation band that extends from 3\,Hz to several kHz. However, it is known that ambient noise in the low-frequency band 10\,Hz -- 20\,Hz at current surface sites of the Virgo and LIGO detectors is predominantly produced by the detector infrastructure. It is of utmost importance to avoid spoiling the quality of an underground site with noisy infrastructure, at least at frequencies where this noise can turn into a detector-sensitivity limitation. In this paper, we characterize the KAGRA underground site to determine the impact of its infrastructure on environmental fields. We find that while excess seismic noise is observed, its contribution in the important band below 20\,Hz is minor preserving the full potential of this site to realize a low-frequency gravitational-wave detector. Moreover, we estimate the Newtonian-noise spectra of surface and underground seismic waves and of the acoustic field inside the caverns. We find that these will likely remain a minor contribution to KAGRA's instrument noise in the foreseeable future.
\end{abstract}

\section{Introduction}
Current gravitational-wave (GW) detectors Virgo \cite{AcEA2015}, LIGO \cite{LSC2015} and KAGRA \cite{AkEA2018} are continuously improving their sensitivities through technology upgrades and commissioning work, and are expected to keep doing so for the next several years \cite{Abbott2020-obs}. However, the plans for Virgo/LIGO detector upgrades do not extend beyond the current decade, and limitations coming from the infrastructures are going to play an increasingly important role. In order to realize the vast science case with state-of-the-art technology \cite{MaEA2020}, new infrastructure is necessary like the proposed Einstein Telescope (ET) \cite{PuEA2010} and Cosmic Explorer (CE) \cite{ReEA2019}. While ET is conceived with the goal to push the low-frequency sensitivity to its limits in a terrestrial environment \cite{HiEA2011}, low-frequency noise poses a greater challenge in CE \cite{HallEA2020}, which is planned as surface infrastructure targeting superior sensitivity at higher frequencies.

Like the KAGRA detector, ET would be underground hosting future-generation detectors with 10\,km long arms. Site quality is of great importance for the construction of the infrastructure as well as for detector operation and science potential \cite{AmEA2020,FiEA2020}. Underground construction helps to strongly reduce the environmental noise (especially that related to seismic and atmospheric sources), but it remains a target of investigations also at KAGRA \cite{washimi2020method,akutsu2021overview}. Especially the terrestrial gravity fluctuations are expected to become an important noise contribution known as Newtonian noise (NN) below 20\,Hz in future detectors \cite{Harms-review}. Newtonian noise is generated by gravity fluctuations associated with density fluctuations of the medium surrounding the test masses. There are several potentially important sources of these fluctuations: seismic fields (body and Rayleigh waves), acoustic fields, and advected temperature and humidity fields \cite{Sau1984, Cre2008}. Another potentially relevant source of NN at KAGRA is underground water. Its amount depends strongly on season, and it needs to be drained, which is done with channels and pipes under the detector arms \cite{Miy2018}.  Modeling of water NN is challenging though, and only upper and lower bounds can be set at the moment \cite{Ken2019}. 

Newtonian noise from the atmosphere can be completely avoided in underground detectors \cite{FiEA2018}, and NN from Rayleigh waves will typically be strongly suppressed underground \cite{Beker2012,BaHa2019}. One of the worries is that detector infrastructure, which includes the ventilation system, pumps, the cryogenic system, and other machines, might cause so much ambient noise that a good part of the advantage of an underground site will be forfeited. In this respect, environmental observations at the KAGRA site provide crucial insight into potential challenges of the design of the ET infrastructure. 

For this purpose, we carried out a characterization of the seismic field at KAGRA with emphasis on the corner station of the detector, where many machines are hosted. We also estimated the NN in the KAGRA detector generated by various types of seismic waves and by the acoustic noise generated by the infrastructure. Results of these analyses are presented in this article. In section \ref{sec:instrumentation}, we summarize the instrumentation and measurements. A basic characterization of seismic fields in terms of their spectra and temporal variations is shown in section \ref{sec:seismic}. Using all available underground seismometers, we infer seismic speeds presented in section \ref{sec:vel-est}. Our estimate of KAGRA seismic and atmospheric NN is reported in section \ref{sec:NN}.

\section{Instrumentation and measurements}
\label{sec:instrumentation}
\begin{figure}
\centering
\includegraphics[width = 1\textwidth]{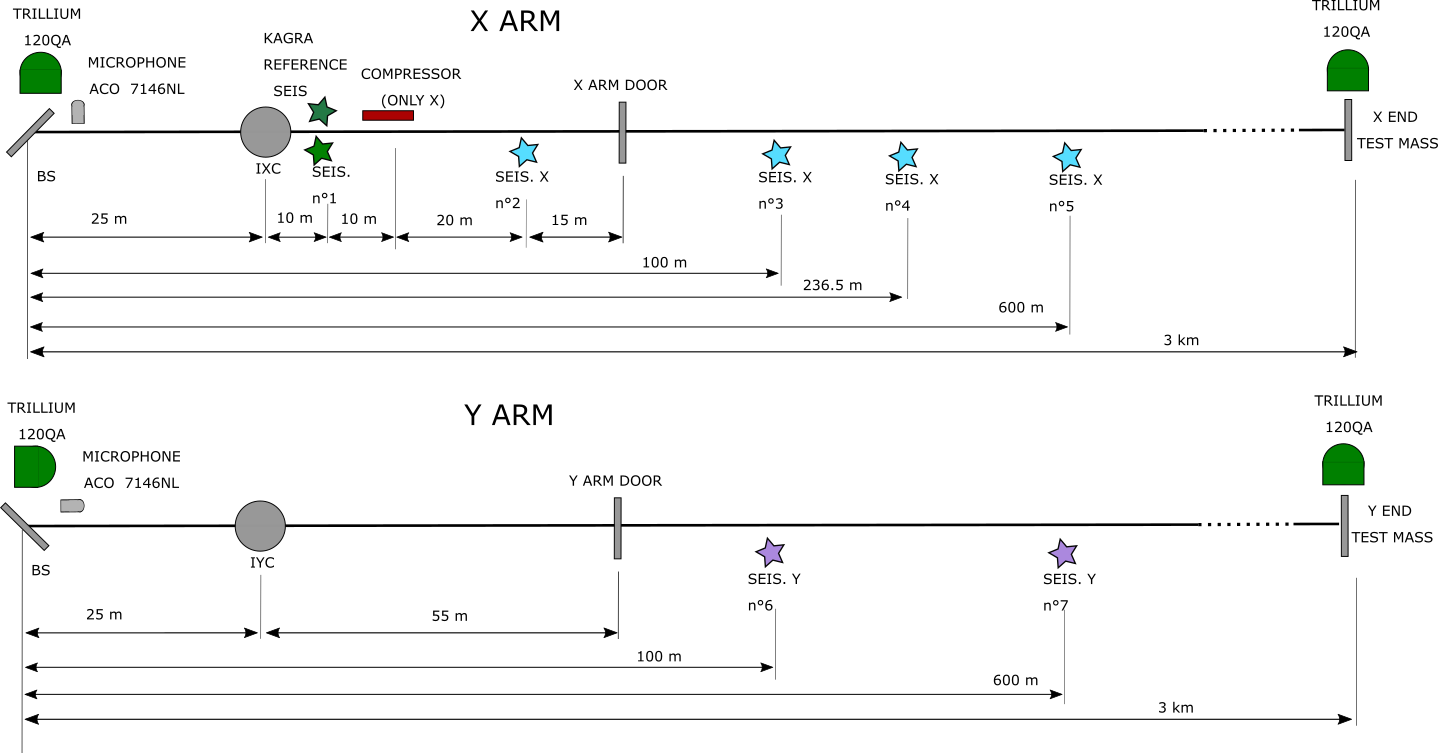}
\caption{Positions where seismic noise was measured moving a Trillium 20s day by day together with a representation of the X arm with the noise sources (red rectangles). \textit{Green star}: KAGRA seismometer used as reference clock.  \textit{Light blue stars}: positions relative to the X arm. \textit{Lilac stars}: positions relative to the Y arm. To be noted that position 1 was not used in the analysis because it served only to synchronize the Trillium 20s with KAGRA's seismic sensors.}
\label{fig:seismometers-positions}
\end{figure}

The KAGRA interferometer's stations (the two end stations and the central one) are hosted at least 200 m below the surface of the Ikenoyama mountain and the bedrock is composed of Hida gneiss \cite{Akutsu2018}.\\  
To investigate the seismic noise generated in the corner station (also called central station, where the beam splitter - BS - and the input test masses are hosted), we used a tri-axial Trillium Compact 20s by Nanometrics \cite{Trillium}, which has flat response to ground velocity between 0.05\,Hz and 100\,Hz. We acquired seismic data at KAGRA's corner station  and along the tunnels of the two interferometer's arms, $X$ and $Y$, each one long 3 km. In this way, we obtained information on how the seismic noise produced in the corner station propagates along the arms and how quickly it attenuates. In \autoref{fig:seismometers-positions}, we draw a scheme to show where we took the seismic measurements. To be noted that position 1 was not used in the analysis because it served only to synchronize the Trillium 20s with KAGRA's seismic sensors.

For what concerns the contribution to the NN caused by body waves, it was estimated using the seismic data from the seismometer location 5, which was farthest from the machines. In this way, we could estimate it in the most quiet environment possible. This was done with the intention of really understanding the NN budget in an underground environment, without additional disturbances.

\begin{figure}
\centering
\includegraphics[width = 0.6\textwidth]{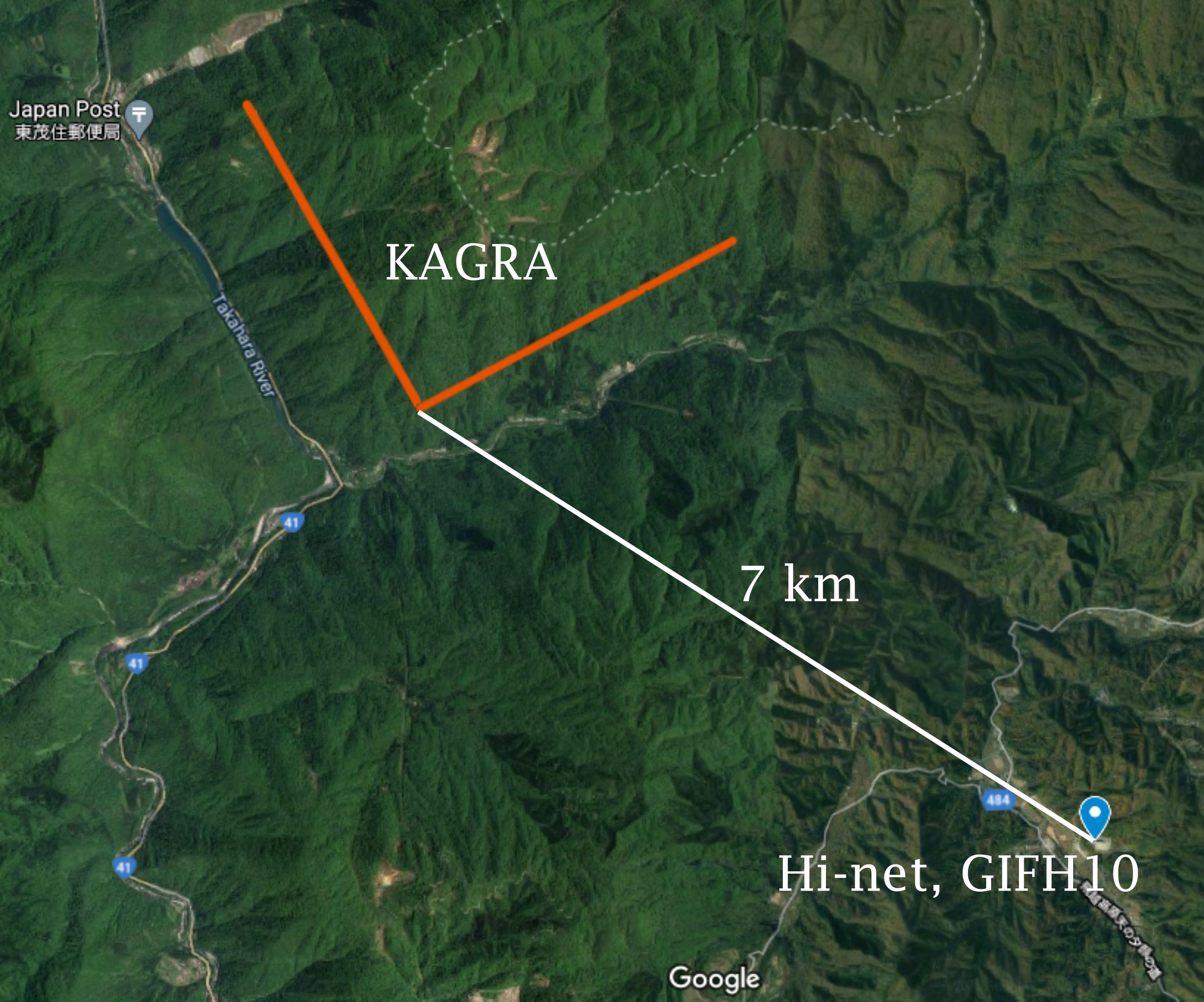}
\caption{Location of the GIFH10 seismic station with respect to the KAGRA detector. The orange lines represent the 3 km long arms of KAGRA.}
\label{fig:SRN-loc}
\end{figure}

The evaluation of the contribution to the NN produced by surface Rayleigh waves was made using the Hi-net data of the GIFH10 station at nearly 7\,km distance to KAGRA (see \autoref{fig:SRN-loc}). It is the closest seismic surface station with publicly available continuous data record. Closer stations of Hi-net and KiK-net only store earthquake triggered time series.  
Given that for the Rayleigh NN estimate we also need the value of the propagation speed of Rayleigh waves, we made a rough estimate of it by means of a simple seismic array already installed at KAGRA. Indeed, KAGRA can synchronize all the equipment by means of a GPS located outside the mine and connected via long cables to all the instrumentation \cite{Akutsu2018}. The seismic array used here was composed of only three sensors (Trillium 120s QA) located in the cavern hosting the BS \cite{akutsu2021overview} and in the two caverns of the end-test masses (see \autoref{fig:seismometers-positions}). We used data collected during 100 quiet periods sampled along one entire year and lasting one hour each. A seismic array composed of three seismometers can only provide limited information about the wavefield, but it is good enough for providing a first, approximate value of the Rayleigh wave speed propagation in order to estimate the related NN.

For our studies, we also use data of a microphone close to the BS \cite{akutsu2021overview}, in the corner station, relatively close (a few tens of meters) to noisy machinery. It is a condenser microphone of model ACO TYPE 7146NL \cite{Mic} with flat response from 1\,Hz to 10k\,Hz, providing high-quality estimates of acoustic spectra in the frequency band of interest.

\section{Seismic noise}
\label{sec:seismic}
\begin{figure}
   	\includegraphics[width=0.9\textwidth]{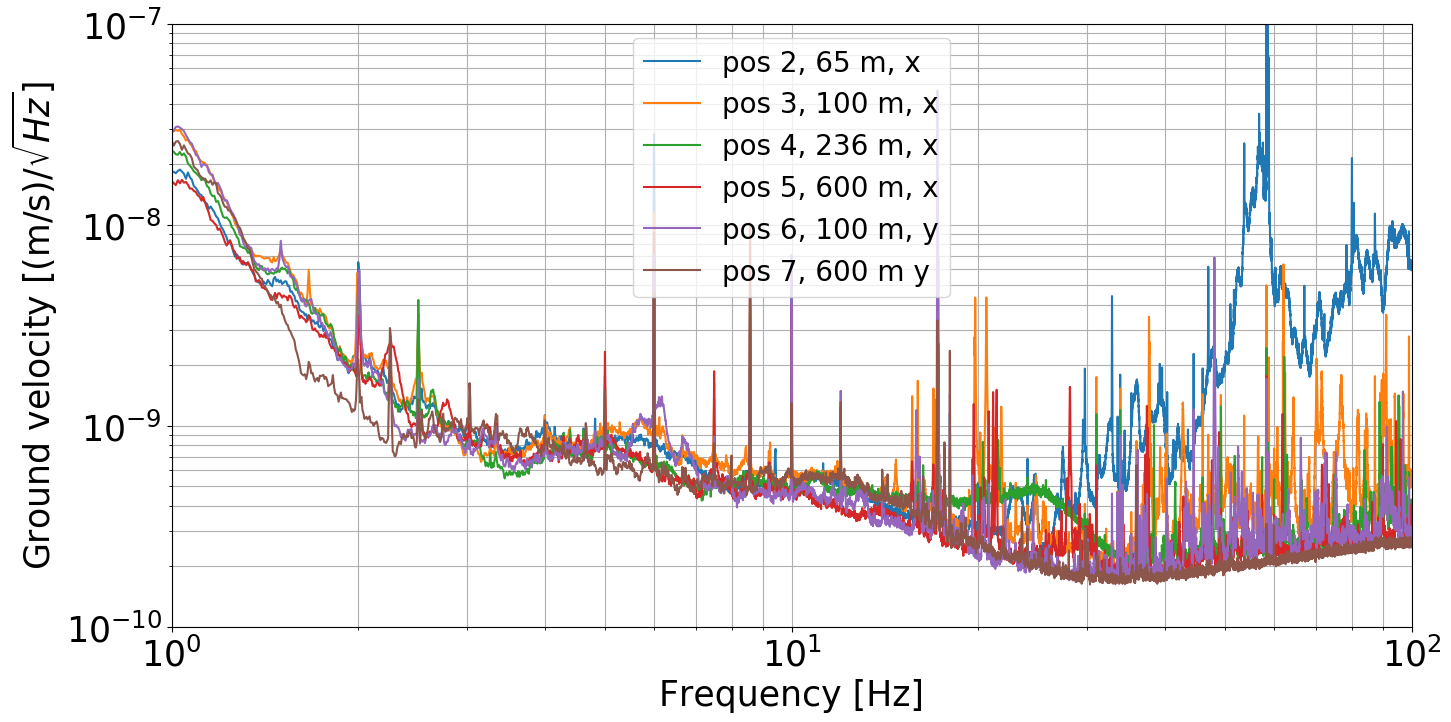}
   	\includegraphics[width=0.9\textwidth]{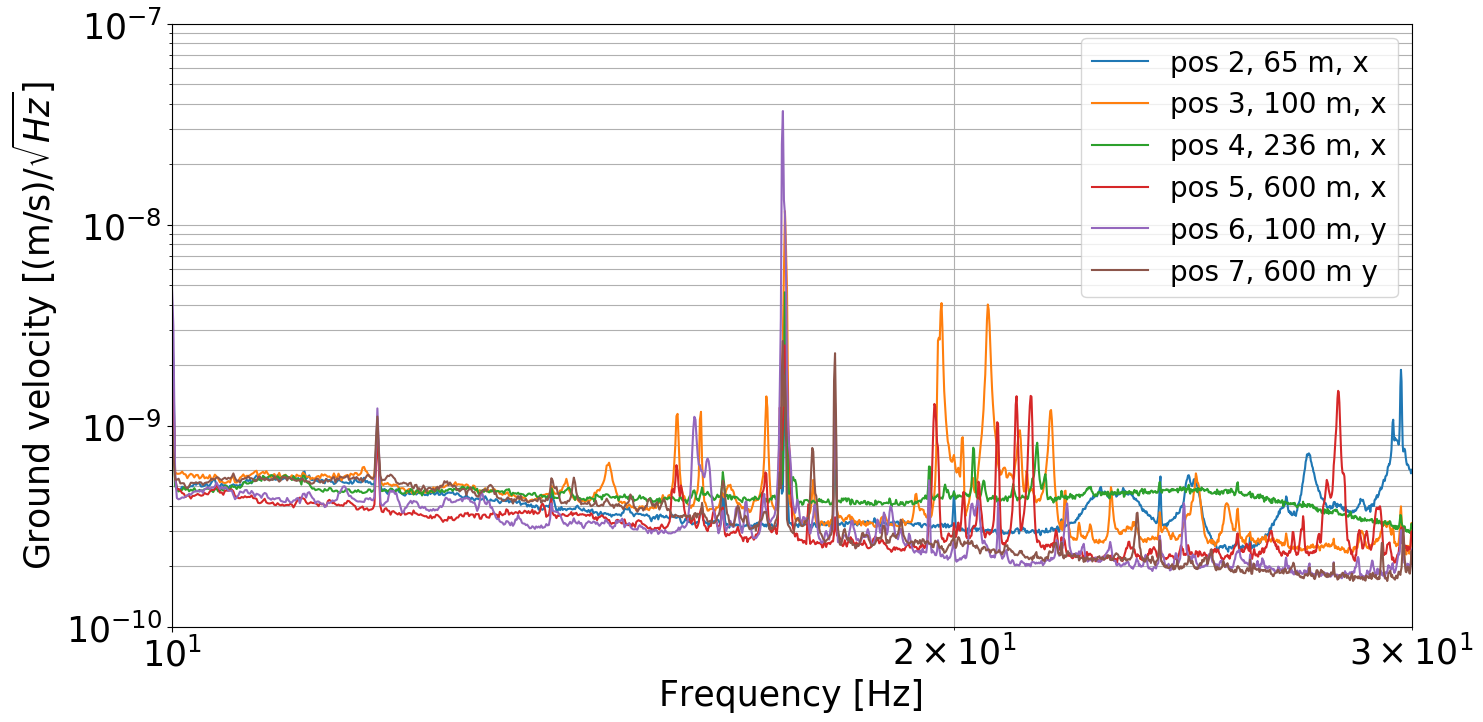}
	\caption{\textit{Top}: spectra of seismic noise as measured along the interferometer arm (see 		\autoref{fig:seismometers-positions} for position references). \textit{Bottom}: zoom taken on 10 -- 30\,Hz.  Distances are with respect to the BS. Data were taken from November 27$^\text{th}$ to December 6$^\text{th}$, 2019. In each selected position, the sensor was deployed on the floor surface and left in acquisition for 1 to 2 days, with a sampling of 250 Hz.  The excess noise observed at position 2 is produced by a compressor, which is switched off during science mode.}
	\label{fig:PSD-seismic}
\end{figure}

In this section, we show the analysis results of the seismic data collected in the locations shown in \autoref{fig:seismometers-positions}. We need to make a remark. The excess noise that we can see in \autoref{fig:PSD-seismic} at position 2, is greatly reduced during GW observations \footnote{http://klog.icrr.u-tokyo.ac.jp/osl/?r=15241}. Indeed, many noisy machinery are switched off during science mode; however, many others will need to run. KAGRA is therefore an important testbed for the ET, indeed, we know that machines such as ion vacuum pumps, cryo-coolers and air conditioners need to work even in science mode to provide the needed working conditions for the detector.\\
The spectra obtained at the different measurement locations are shown in \autoref{fig:PSD-seismic}. At position 2, excess noise in the range 20\,Hz -- 100\,Hz can be seen. It is produced by a compressor used for the operation of a clean room, and it is switched off during science mode. It rapidly attenuates leaving no clear sign of excess noise at position 3 just 35\,m away from position 2. A possible explanation is that acoustic noise produced by machines causes localized vibrations of the ground, or acoustic noise could directly act on the seismometer causing vibrations of its frame. 

These results have important implications for infrastructure noise in underground environments. It seems that machines necessary for the operation of KAGRA, which includes the cryogenic and ventilation system, do not generate significant seismic disturbances. The only important effect that seems to happen is direct forcing of acoustic fields on the ground, which are very localized. This is in stark contrast to our experience at the LIGO and Virgo sites, where infrastructure noise is known to propagate in the form of seismic waves and dominating fully or partially the seismic spectra \cite{Tringali2019,HaEA2020,Badaracco2020,FiEA2020}. Given that the natural seismic background noise is extremely low at the KAGRA site, it was not to be expected that machines had so little influence on the seismic environment. While the reasons are not known yet, a possible explanation is that the machines are mounted on very stiff rock, while machine at surface sites are located on concrete slabs of 1\,m -- 2\,m thickness with soil layers underneath. In the latter case, it is easier to couple vibrational energy into seismic waves. Also, it might be that machines are simply less noisy at the KAGRA site. This needs to be explored in the future.

\begin{figure}[h!]
\begin{minipage}{1\textwidth}
   	\includegraphics[width=0.5\textwidth]{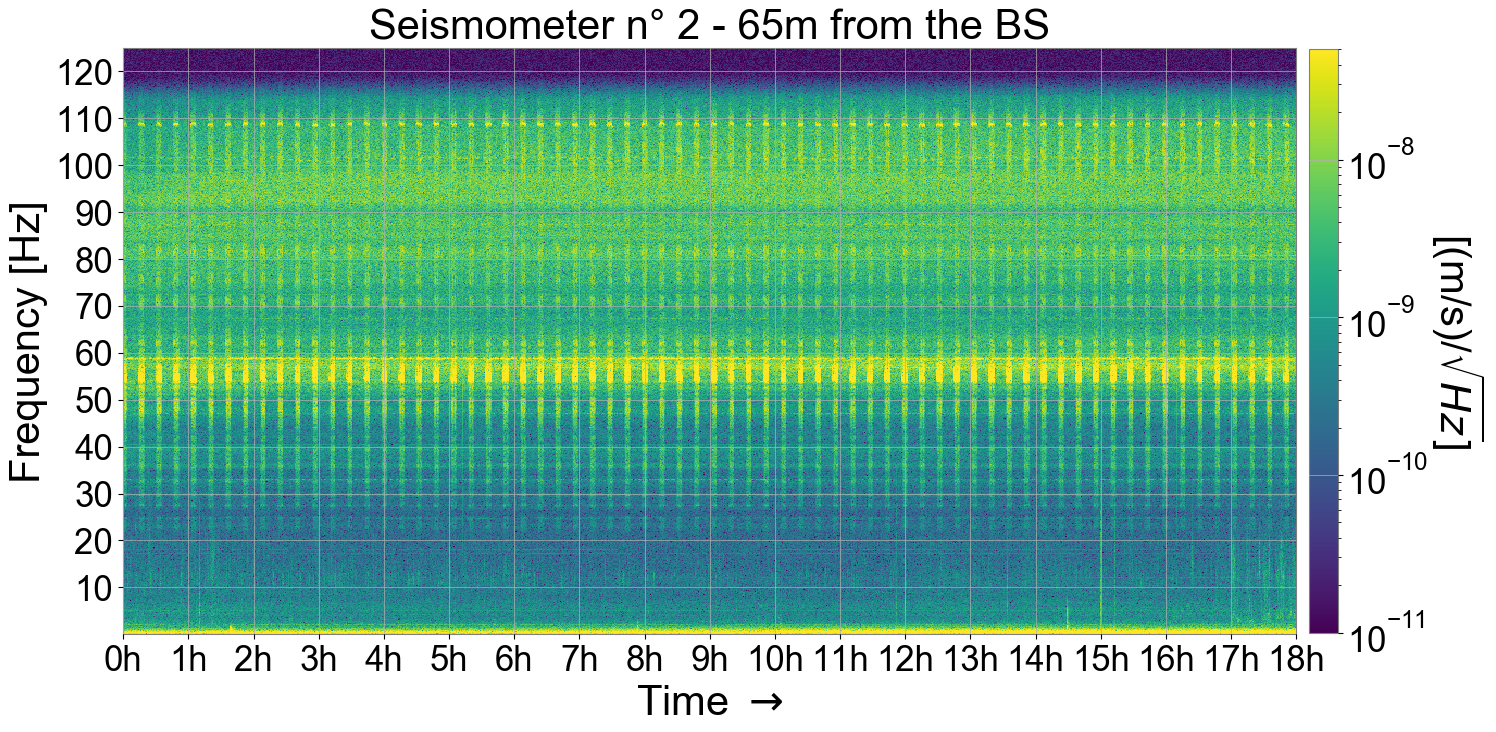}
    \includegraphics[width=0.5\textwidth]{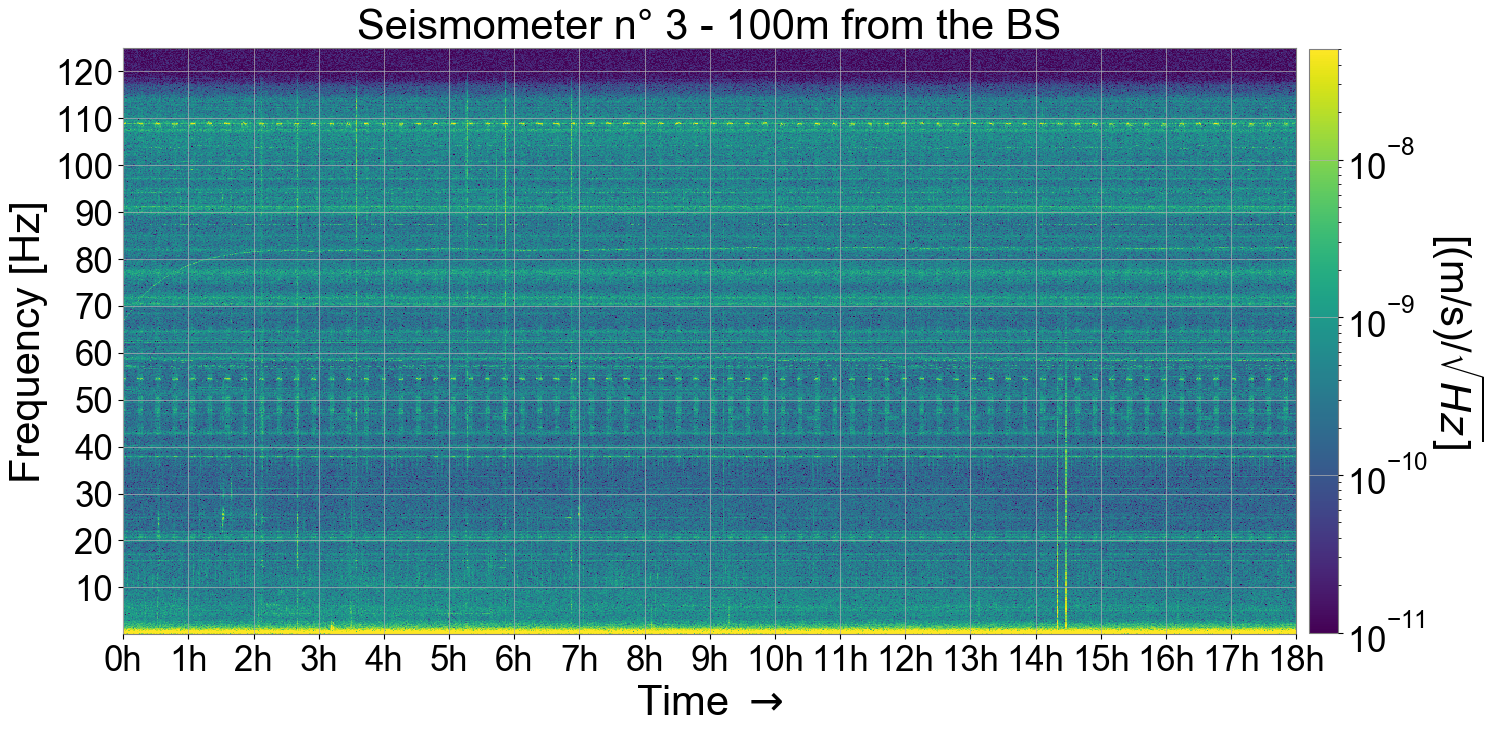}
   	\vfill
   	\includegraphics[width=0.5\textwidth]{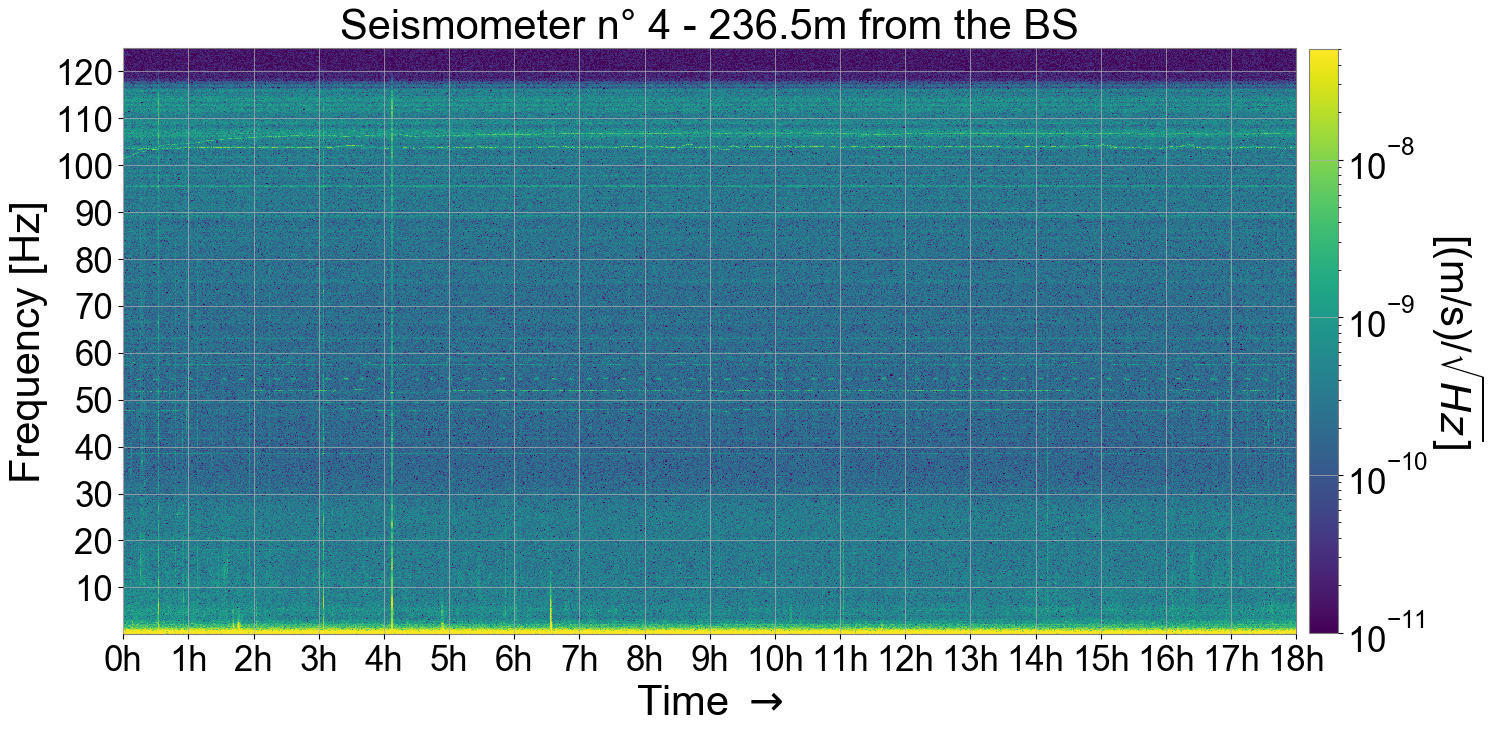}
    \includegraphics[width=0.5\textwidth]{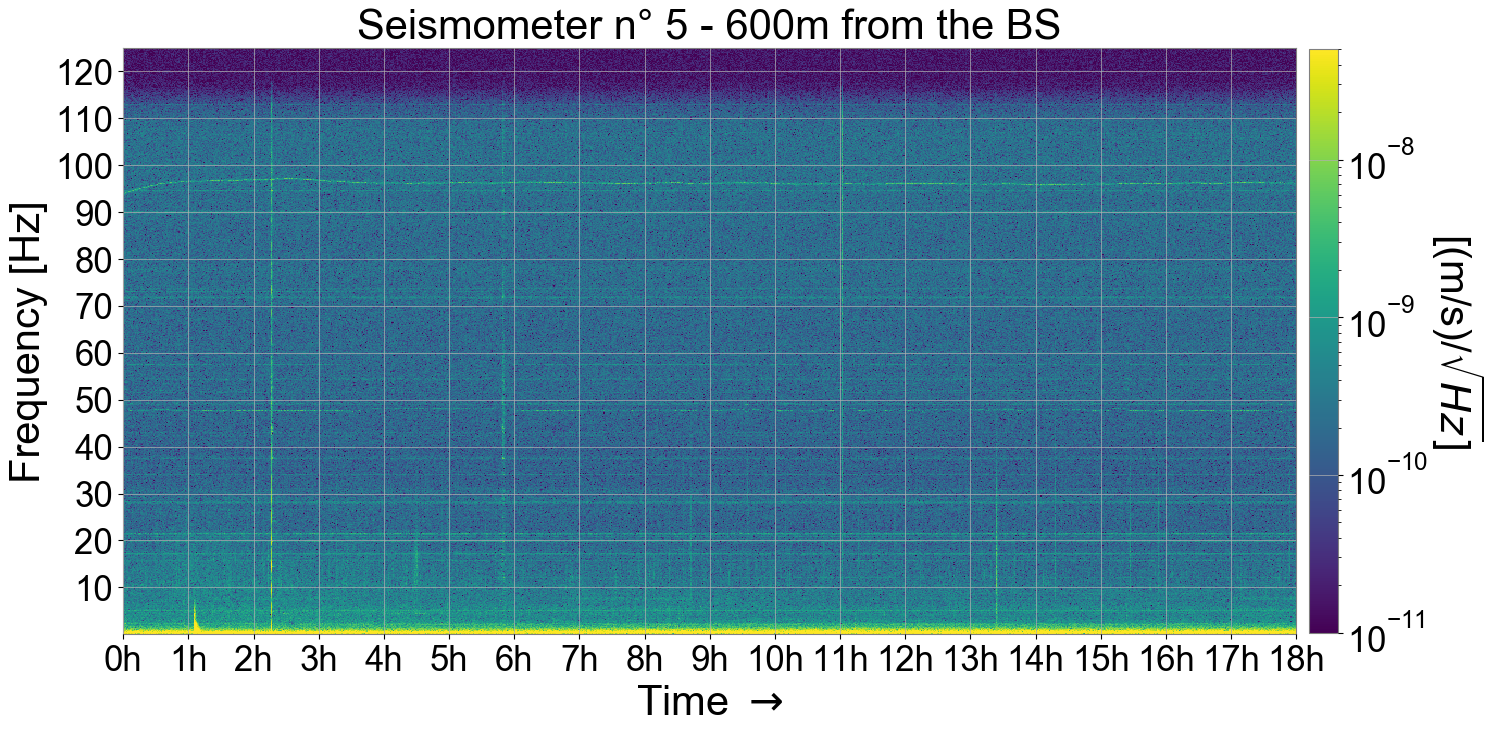}
    \vfill
   	\includegraphics[width=0.5\textwidth]{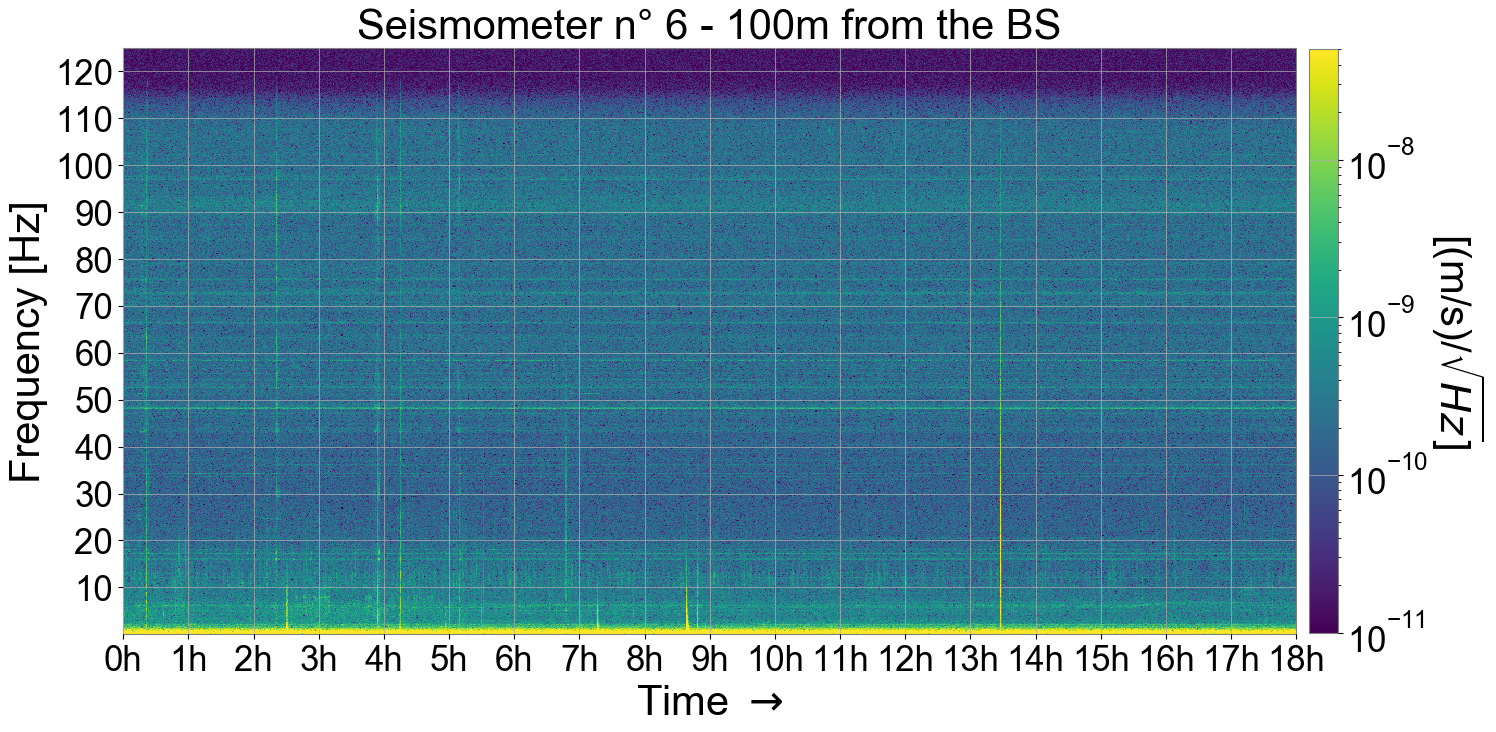}
    \includegraphics[width=0.5\textwidth]{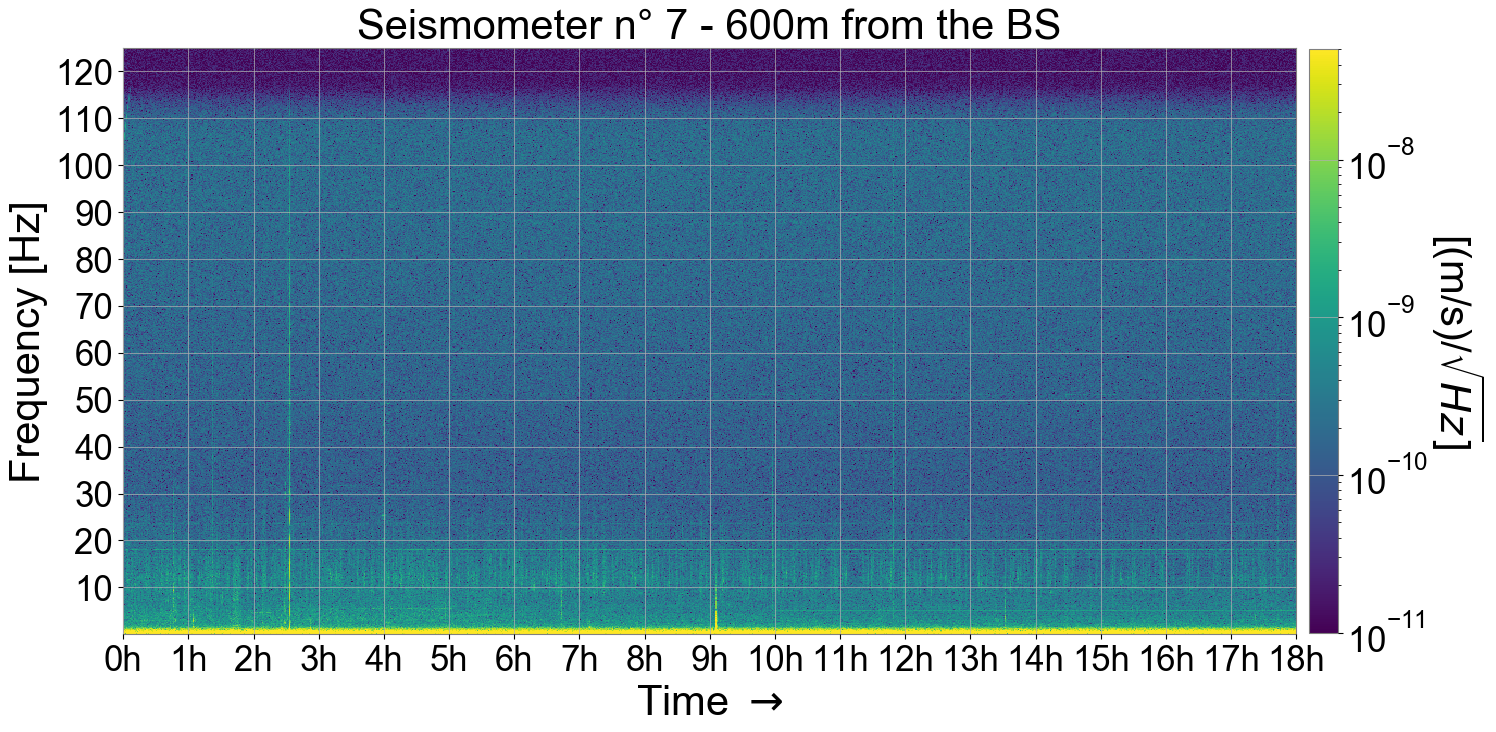}
	\vfill
\caption{Seismograms relative to the different locations at which the Trillium was set (see \autoref{fig:seismometers-positions}). In position 2 the presence of noise due to the compressor is clearly visible. Each plot was made taking 18 consecutive hours of data. Position 2 starts from 6:00 a.m. on 27/11/2019, Position 3 starts from 2:00 a.m. on 02/12/2019, Position 4 starts from 6:00 a.m. on 28/11/2019, Position 5 starts from 3:00 a.m. on 29/11/2019, Position 6 starts from 2:00 a.m. on 04/12/2019, Position 7 starts from 2:00 a.m. on 03/12/2019.}
\label{fig:spectrograms}
\end{minipage}
\end{figure}

In \autoref{fig:spectrograms}, spectrograms of the seismic noise are shown for the six positions where the seismometer was placed. Also here, we can see that the noise generated inside the corner station decreases rapidly as we move away from the station along the two arms. Furthermore, especially the spectrogram at position 7 indicates a substantially increased level of stationarity of the seismic field (as an example, at 30\,Hz the root mean square of the spectrum at position 7 is a factor 5 smaller than that at position 2 and a factor 3 smaller than that of position 6). 
The study of seismic transients associated with seismic waves and not with an acoustic field acting on the ground gives us another important piece of information. The observations indicate that even if a disturbance occurs that produces seismic waves, then at a distance $>500\,$\,m the seismic amplitudes has decreased significantly hardly showing in spectrograms. This can serve as a first model for a better positioning of potentially noisy machines relative to test masses (even though we have also seen that the machines do not seem to produce seismic waves with significant amplitude).

\begin{figure}[h!]
\begin{minipage}{1\textwidth}
   	\includegraphics[width=0.5\textwidth]{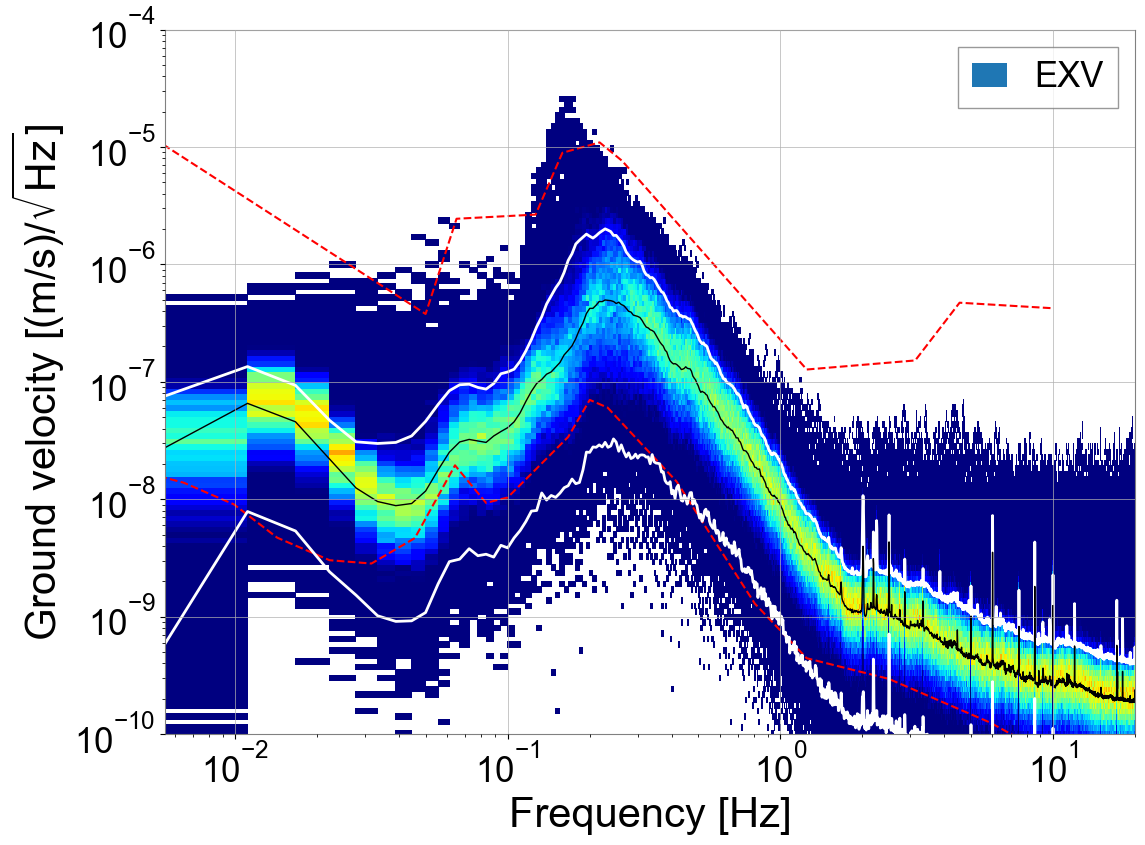}
    \includegraphics[width=0.5\textwidth]{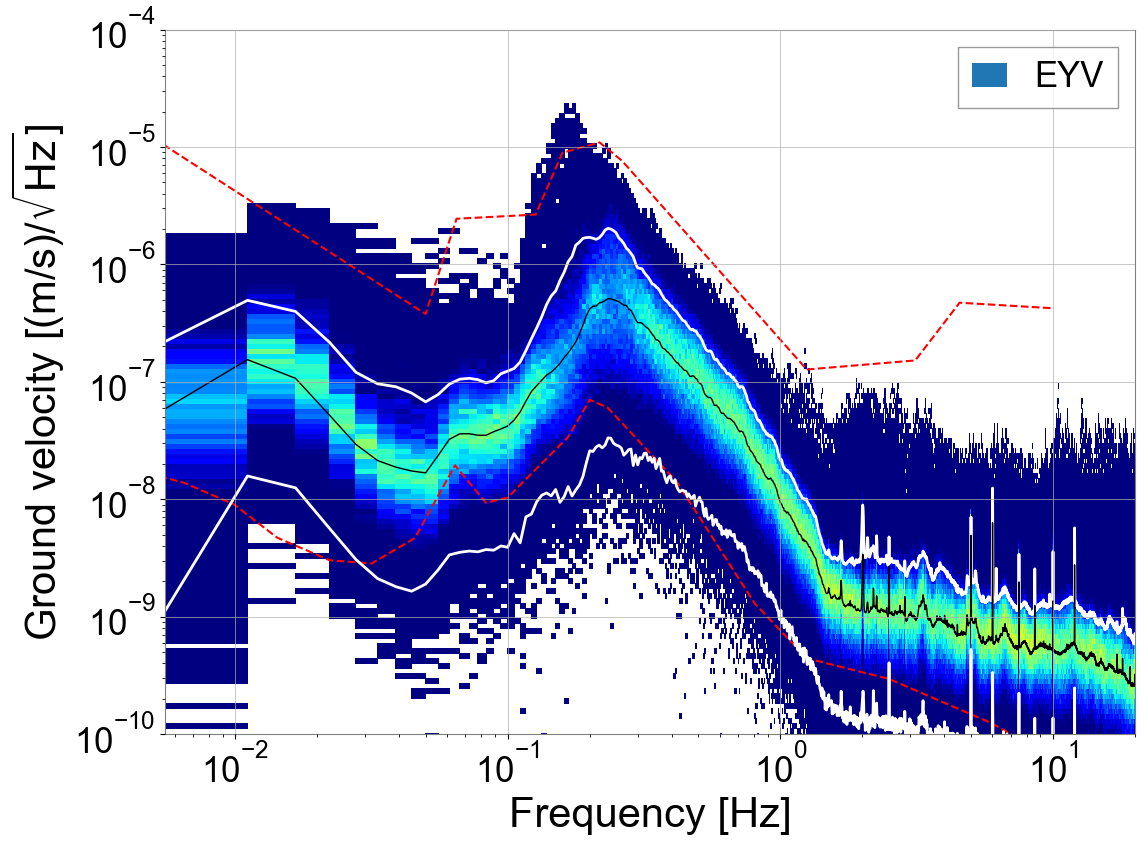}
   	\vfill
   	\centering
   	\includegraphics[width=0.5\textwidth]{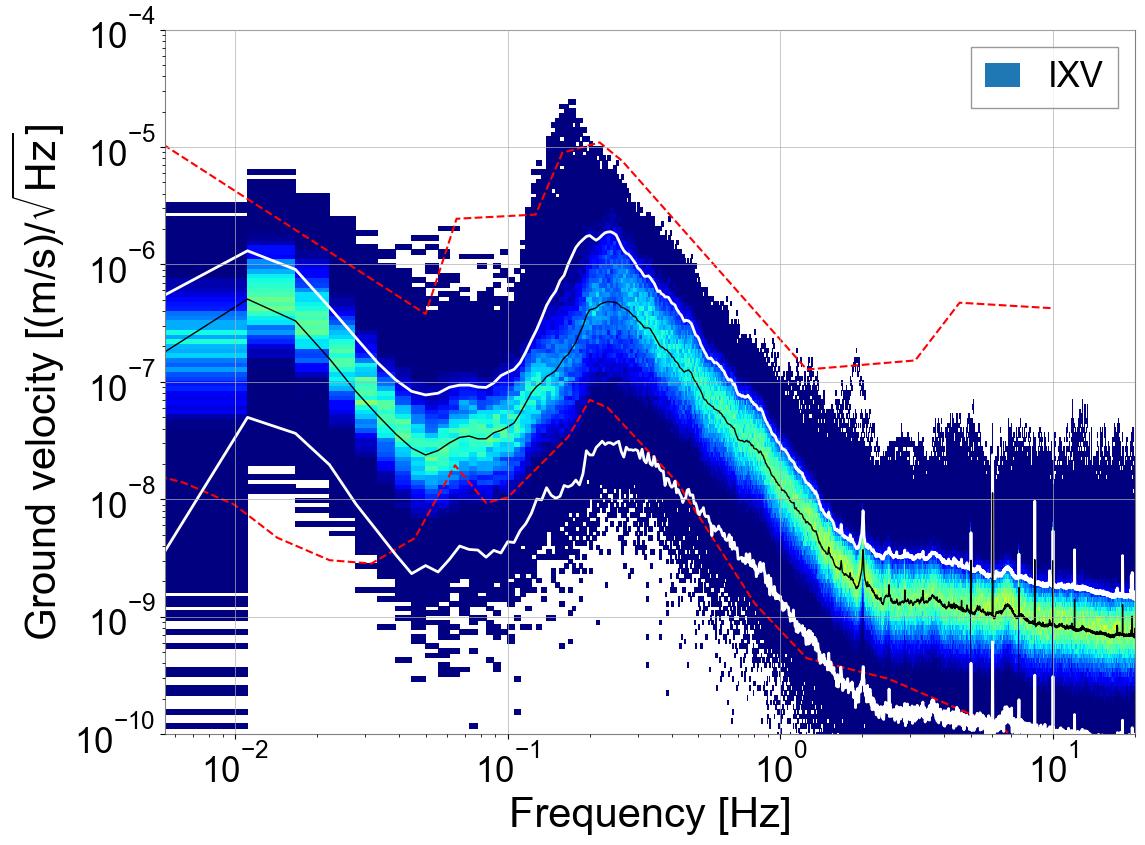}
   	\caption{We show here their spectral variations (over a year of quiet times) together with the global low-noise and high-noise models (red dashed lines) \cite{Pet1993} and the 10$^{\rm th}$, 90$^{\rm th}$ (white lines) and 50$^{\rm th}$ (black line) percentiles. \textit{Upper left}: X-end mass upper cavern; \textit{upper right}: Y-end mass upper cavern; \textit{bottom}: X-input mass upper cavern.}
	\label{fig:PSD-exv-ixv-eyv}
\end{minipage}
\end{figure}
Test masses in KAGRA are suspended from chains of mechanical filters and pendula that reach about 13\,m up to separate caverns built above the main interferometer caverns \cite{Michimura_2017}. Seismometers are deployed inside these caverns monitoring seismic motion close to the where the first stage of the isolation system is mounted. Histograms of spectra of vertical ground motion at the two end-test masses (EXV, EYV) and the input-test mass (IXV) of the X-arm are shown in \autoref{fig:PSD-exv-ixv-eyv}. At all three stations, seismic spectra are exceptionally quiet between 1\,Hz and 20\,Hz, and the end stations are a bit quieter than the corner station IXV. The histograms show that there are occasional outliers adding to the stationary background increasing the higher percentiles of the histograms. However, the main mode of the histograms in the range 1\,Hz to 20\,Hz is very clear indicating a stable natural background.

\section{Velocity estimation in KAGRA} 
\label{sec:vel-est}
\begin{figure}[h!]
\begin{minipage}{1\textwidth}
    \includegraphics[width=0.5\textwidth]{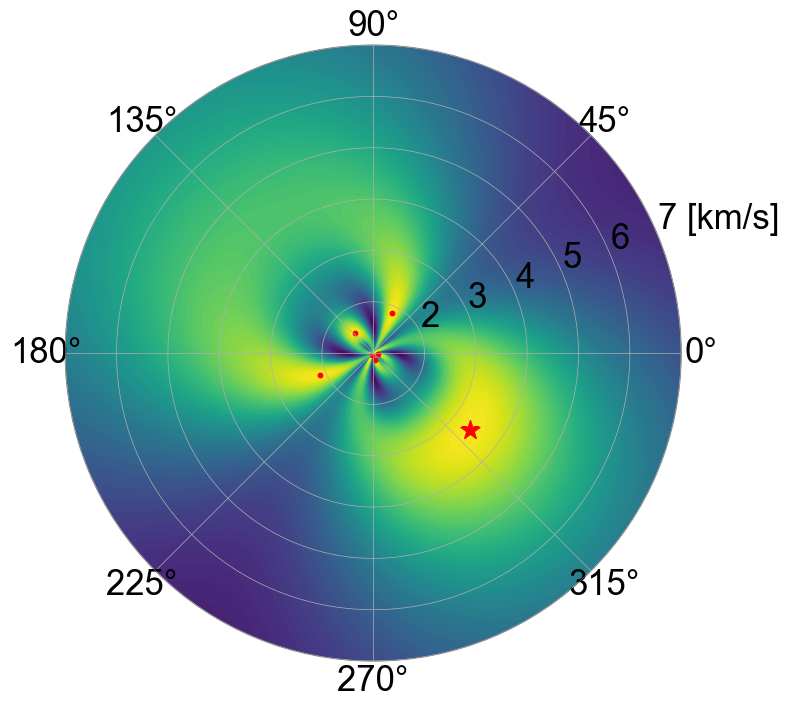}
    \includegraphics[width=0.5\textwidth]{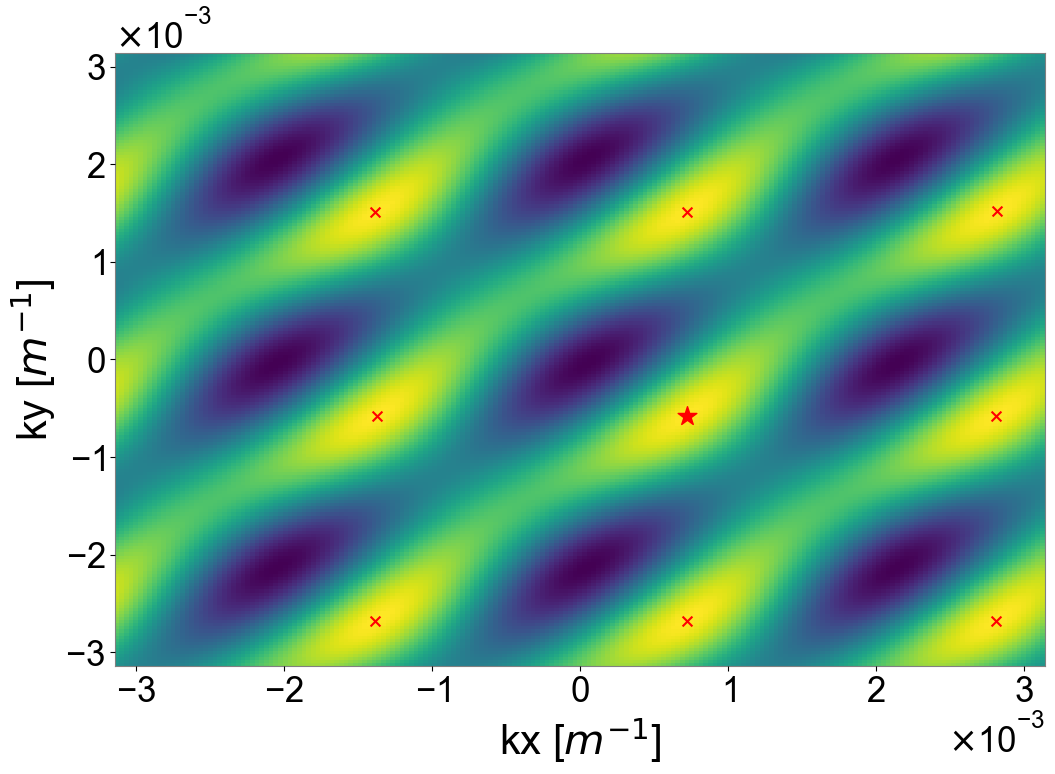}
    \caption{The 9 peaks of \autoref{eq:beamforming} in the portion of the wave vector space analysed at 0.5\,Hz. They are represented both in the polar plane (\textit{left}) and in the Cartesian plane (\textit{right}).}
    \label{fig:array-response+aliasing}
\end{minipage}
\end{figure}

In this section, it will be explained how the velocities of Rayleigh waves were estimated using the small array already described in \autoref{sec:instrumentation}. To extract the velocity values we used the Bartlett beamforming technique \cite{Krim1996}. Its aim is to find the values of the wave vector, \textbf{k}, that maximize the output power of the array defined as:
\begin{equation}\label{eq:beamforming}
P_Y(\mathbf{k}) = \frac{\mathbf{w}^\dag(\mathbf{k})\cdot\mathbf{R}\cdot\mathbf{w}(\mathbf{k})}{\mathbf{w}^\dag(\mathbf{k})\cdot\mathbf{w}(\mathbf{k})},
\end{equation} 
where $\mathbf R$ is the correlation matrix between seismometers estimated from measurements, and $\mathbf w$ is a weighting vector. The shortest distance between the three seismometers used for our studies is $L = 3$\,km. This gives us a very good resolution in wave-vector space, $\Delta k= \pi/L \sim 10^{-3}$\,m$^{-1}$, but the distance is too large for straight-forward analyses in the NN band between 5\,Hz and 20\,Hz. Aliases of the physical mode can be seen in the spatial spectrum, and it is not possible without additional information to selected between the modes. An example spatial spectrum obtained with data from the three seismometers is shown in \autoref{fig:array-response+aliasing}. 

\begin{figure}[h!]
	\centering
	\begin{minipage}{1\textwidth}
		\centering
	    \includegraphics[width=1\textwidth]{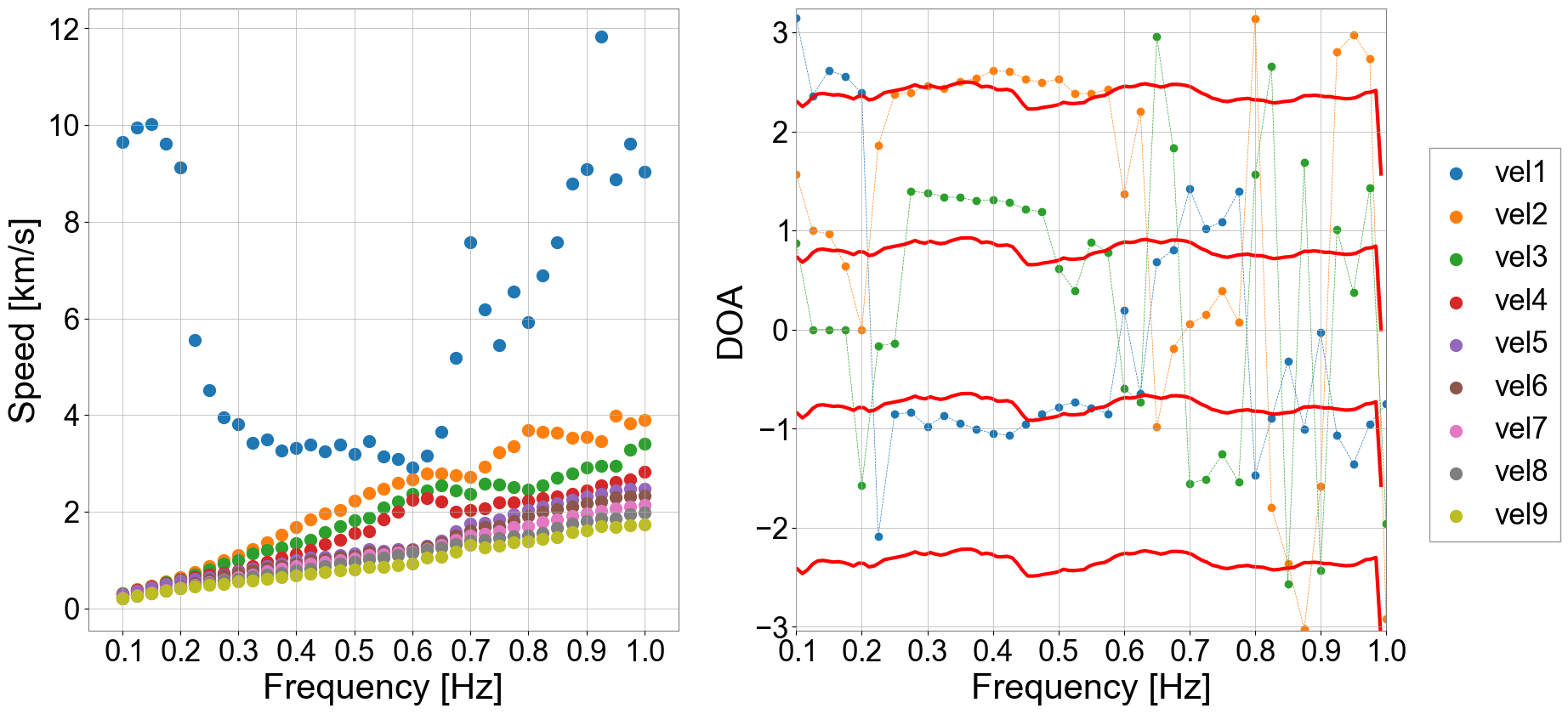}
    \caption{Values of the 9 median ranked speeds (\textit{left}). Comparison of the DOA from the first three peaks (blue: first peak, orange: second peak and green: third peak) with the DOA inferred from the horizontal channels (red lines). See in the text to further explanations (\textit{right}).}
    \label{fig:KAGRA-velocities-DOA}
\end{minipage}
\end{figure}

We searched for all the peaks contained in the wave vector space with $k_x, k_y \in [-3\Delta k, 3\Delta k]$ and then we ranked them by speed values in descending order. The wave-number range allows us to analyze waves with speeds down to $v_{\rm min}\sim 2\,{\rm km}\cdot f$. For a given frequency, we searched for the $\mathbf{k}$ that maximized \autoref{eq:beamforming}, and this was done for each one of the 100 data segments. In \autoref{fig:KAGRA-velocities-DOA}, we plot the median values found for the 9 ranked speeds (in the analysed portion of k-space, $[-3\Delta k, 3\Delta k]$, there is space for 9 peaks, so 9 velocities). If we look at the median velocity corresponding to the fastest mode (the one closest to the origin in k-space), we can see that above 0.6\,Hz, there is an increase in the velocity values. This is very likely due the fact that the fastest peak does not correspond anymore to the physical one, and is now an alias of a slower physical mode. Looking at the median values of the speeds of other ranked peaks, we notice that above 0.6\,Hz, the median values obtained with the peaks from 4 to 9 follow a linear trend: this suggests that they are the result of aliasing. Indeed, an aliased mode can be written as:
\begin{equation}
k_{x,y} = k_{0_{x,y}} + \frac{2\pi N}{\Delta_{x,y}}
\end{equation}
where $\Delta_{x,y}$ is the largest distance along the $x$ or $y$ direction, $k_{0_{x,y}}$ is the physical mode that we are searching for and $k_{x,y}$ is the $x$ or $y$ component of the peak located in $\mathbf{k}$. N is an integer number.  This means that the velocity is 
\begin{equation}
v = \frac{\omega}{\sqrt{k_x^2+k_y^2}} = \frac{\omega}{\sqrt{(k_{0_x}+2\pi N/L)^2+(k_{0_y}+2\pi N/L)^2}}
\end{equation} 
and if $2\pi N/L$ is large compared to $k_{0_{x,y}}$ the speeds will have a linear trend with respect to the frequency $\omega$: $v = \omega L/(2\pi N)$. We discard speeds obtained from peaks 4 to 9. After 0.6\,Hz, we need to understand which peak between 2 and 3 is the correct one. For this, we can look at the plot of the directions of arrival (DOA) (\autoref{fig:KAGRA-velocities-DOA}) that were calculated from the first three ranked peaks (blue, green, orange). We notice that below 0.6\,Hz, the DOAs are more uniform suggesting a single dominant ocean-wave field producing the microseisms in this band. Above 0.6\,Hz, the DOAs vary more strongly between frequencies due to a more complex composition of the seismic field emitted from several sources.

For the analyses, we used the vertical displacement, which should be dominated by Rayleigh waves below 1\,Hz even in the depth of Mount Ikenoyama. Rayleigh waves have an elliptical polarization composed of a vertical component and a horizontal component along the direction of propagation \cite{AkRi2009}. Therefore, the propagation direction of a wave can also be determined by observing the displacement in horizontal direction. For this purpose, we calculated the PSD of the horizontal channels along $x$ and $y$ and identified the propagation directions consistent with the two PSD values (there are always 4 such directions). In the right plot of \autoref{fig:KAGRA-velocities-DOA}, we can see the four directions (red) consistent with the horizontal displacements. Above 0.6\,Hz, excluding the DOA from the first peak, we can see that the one from the third peak better follows the red lines, so we are led to believe that the speeds from the third peak correspond to physical values. 

\begin{figure}[h!]
	\centering
	\begin{minipage}{1\textwidth}
		\centering
	    \includegraphics[width=0.65\textwidth]{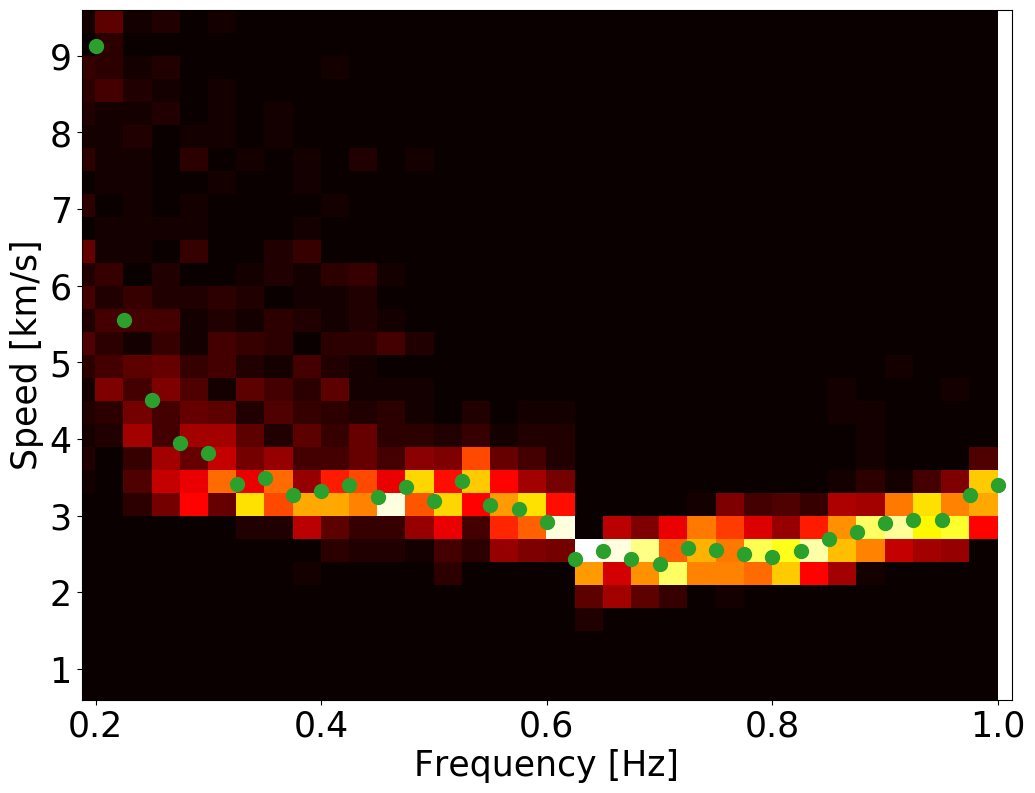}
    \caption{Histogram of seismic speeds given by the first ranked peak (up to 0.6\,Hz) and by the third one (from 0.6\,Hz to 1\,Hz).}
    \label{fig:KAGRA-velocities}
\end{minipage}
\end{figure}

In \autoref{fig:KAGRA-velocities}, we show the histogram of median speeds obtained using the first peak (up to 0.6\,Hz) and the third one (beyond 0.6\,Hz). Analyses below 0.2\,Hz are not possible since the spectral resolution is not sufficient anymore to resolve propagation directions.
We can see that from 0.2\,Hz to 0.6\,Hz velocities diminish with increasing frequency. This is in agreement with some other studies where we can see this trend \cite{Nishida2008}. Values slightly above 3\,km/s are reasonable since we can expect that the microseism below 1\,Hz consists mainly of fundamental Rayleigh waves \cite{BonnefoyClaudet2006}. We can also see that, if we consider a depth of $\sim 200$\,m for the corner station \cite{Akutsu2018}, at 0.5\,Hz and with a seismic velocity $v = 3$\,km/s (\autoref{fig:KAGRA-velocities}), the attenuation factor of the Rayleigh waves is only $e^{-hk} \sim 0.8$, so it is reasonable to assume that Rayleigh waves at frequencies $<1$\,Hz are also present underground at KAGRA's corner station.

\section{NN estimate in KAGRA}
\label{sec:NN}
\begin{figure}[h!]
\centering
\begin{minipage}{1\textwidth}
   	\includegraphics[width=0.9\textwidth]{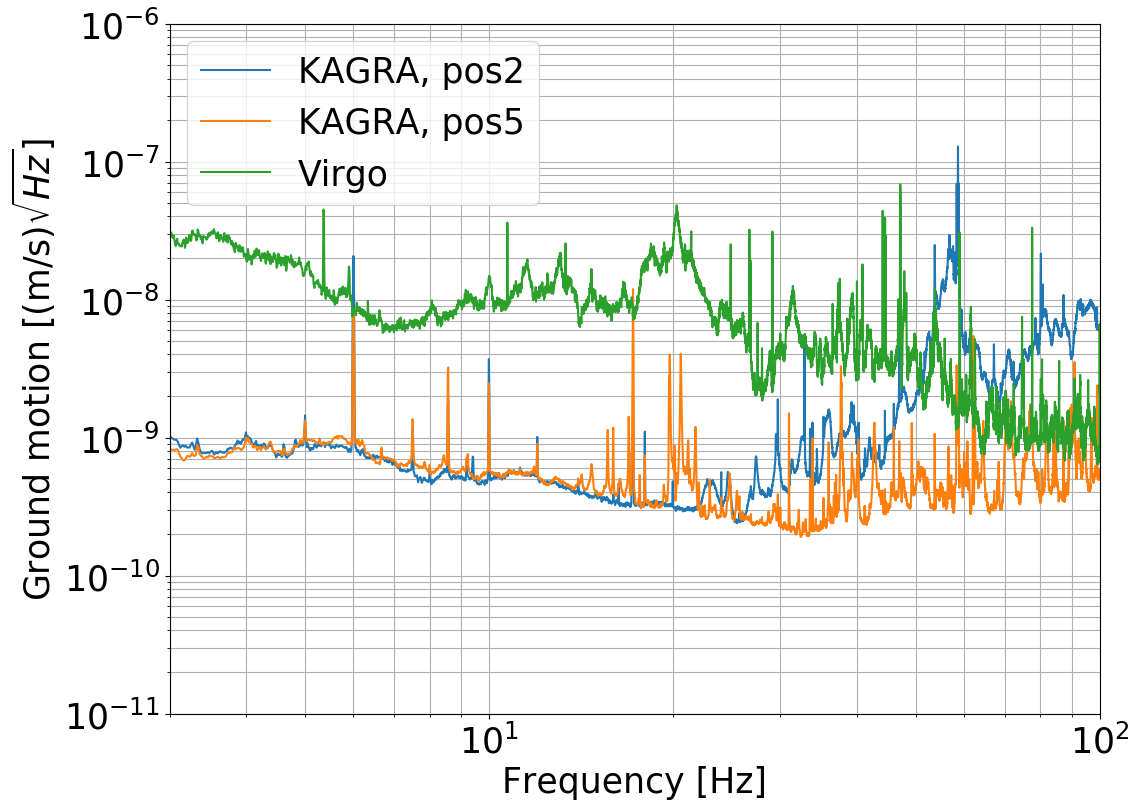}
    \caption{Comparison between seismic spectra at KAGRA (position 2 and 5 of \autoref{fig:seismometers-positions}) and Virgo.}
    \label{fig:KAGRA-Virgo}
\end{minipage}
\end{figure}

KAGRA was constructed underground since certain advantages are expected for the operation of the interferometer and for its sensitivity. Newtonian noise from seismic and atmospheric fields are reduced. The reduced seismic vibrations also make it easier to control the interferometer, which can increase duty cycle, but also reduce control noise for another potential gain in sensitivity. We can see in \autoref{fig:KAGRA-Virgo} that at 10\,Hz the measured seismic spectrum at KAGRA is 2 orders of magnitude smaller than the one measured at Virgo. 

Surface waves are characterized by an exponential decay $\exp(-h\omega/v)$ of their amplitude, where $h$ is the depth, $v$ a speed parameter of Rayleigh waves (there are two such parameters and both are different from the horizontal propagation speed \cite{AkRi2009}). This suppression is more effective towards higher frequencies. Going underground helps to greatly reduce the surface contribution to NN, but the NN from the seismic body waves remains. It is then important to understand how much NN is still present in underground detectors. A first estimate of seismic NN at KAGRA was presented in \cite{Somiya_2012}, where a few educated guesses had to be done for certain geophysical parameters at the KAGRA site. Here, we refine the model using the seismic analyses presented in the previous sections.

To estimate the body-wave NN budget in the NN band, we can use the data taken inside the arm's tunnel, in particular, we used the data taken in at position 5; see \autoref{fig:seismometers-positions}. The following model is employed \cite{Harms-review}:
\begin{equation}\label{eq:acc-P}
S\left(\delta a^P; \omega\right) = \left(\frac{8}{3}\pi G \rho_0\right)^2S\left(\xi^P; \omega\right),
\end{equation}
which is conservative as it assumes the worst-case scenario where only compressional waves (P-waves) are present (shear-wave content reduces body-wave NN). This estimate is strictly valid only in an infinite and homogeneous medium, but since a major contribution to this noise comes from the displacement of cavern walls, it is relatively insensitive to geology and detector depth (in contrast to Rayleigh NN, which depends strongly on both). Here, $S\left(\delta a^P; \omega\right)$ represents the PSD of the NN acceleration of the test mass provoked by the seismic displacement along the direction of the interferometer arm with PSD $S\left(\xi^P; \omega\right)$. Assuming that the noise is the same at all 4 test masses and uncorrelated between them, the test-mass acceleration noise must be multiplied by $2/(L\omega^2)$, $L$ being the length of a detector arm, to obtain the associated strain noise. We also point out that the exact geometry and radius of the caverns only weakly influence NN as long as the seismic waves are much longer than the cavern diameter \cite{Harms-review}.
\begin{figure}[h!]
	\begin{minipage}{1\textwidth}
 	  	\includegraphics[width=0.5\textwidth]{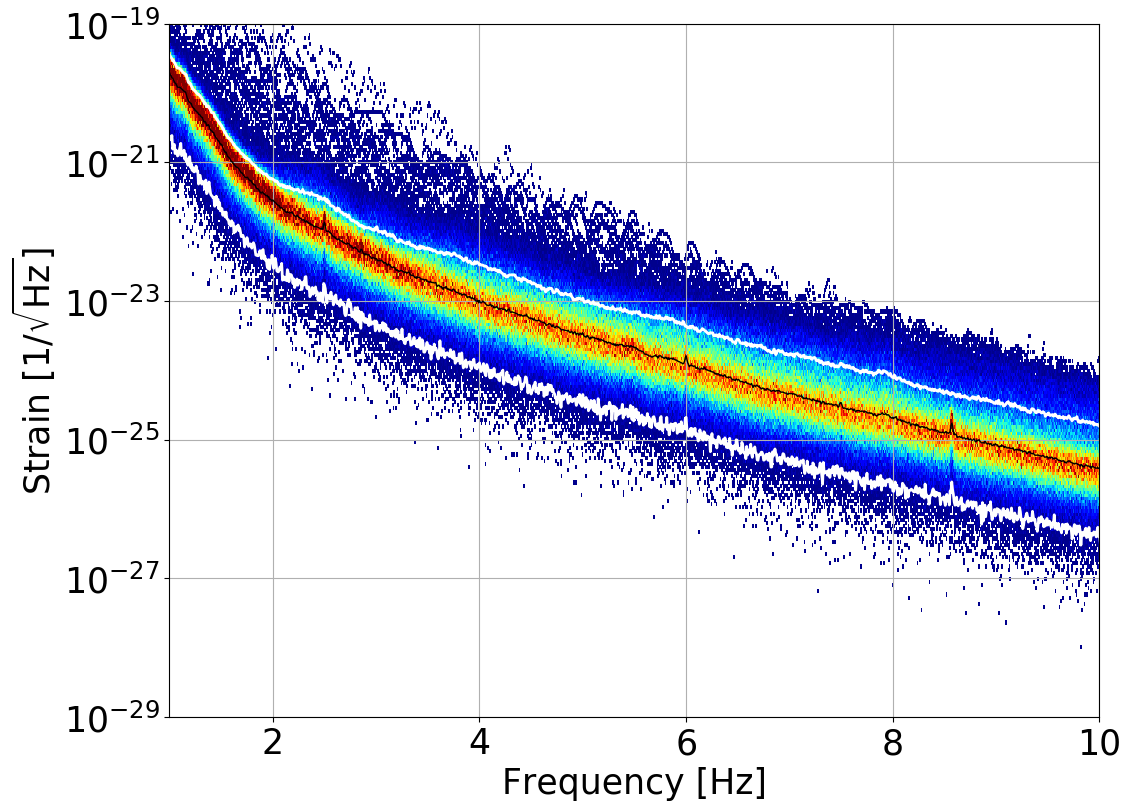}
 	   \includegraphics[width=0.5\textwidth]{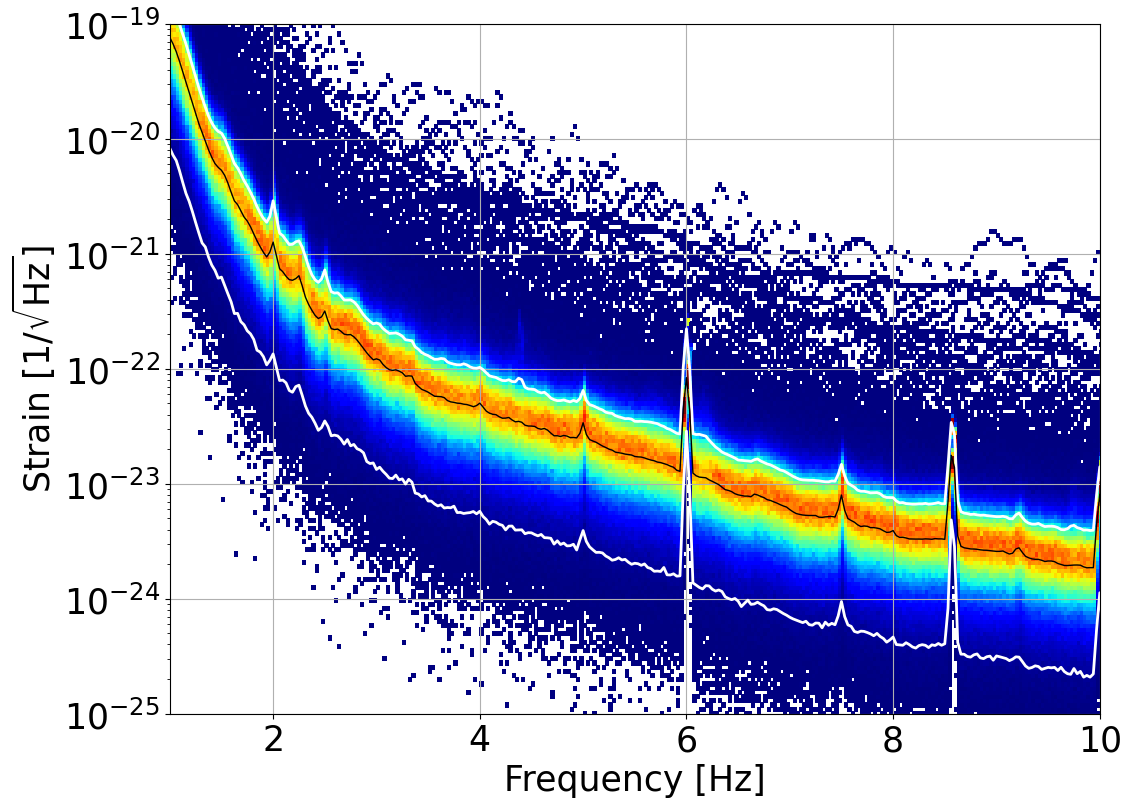}
 	 \end{minipage}
	\begin{minipage}{1\textwidth}
	\centering
 	   \includegraphics[width=0.5\textwidth]{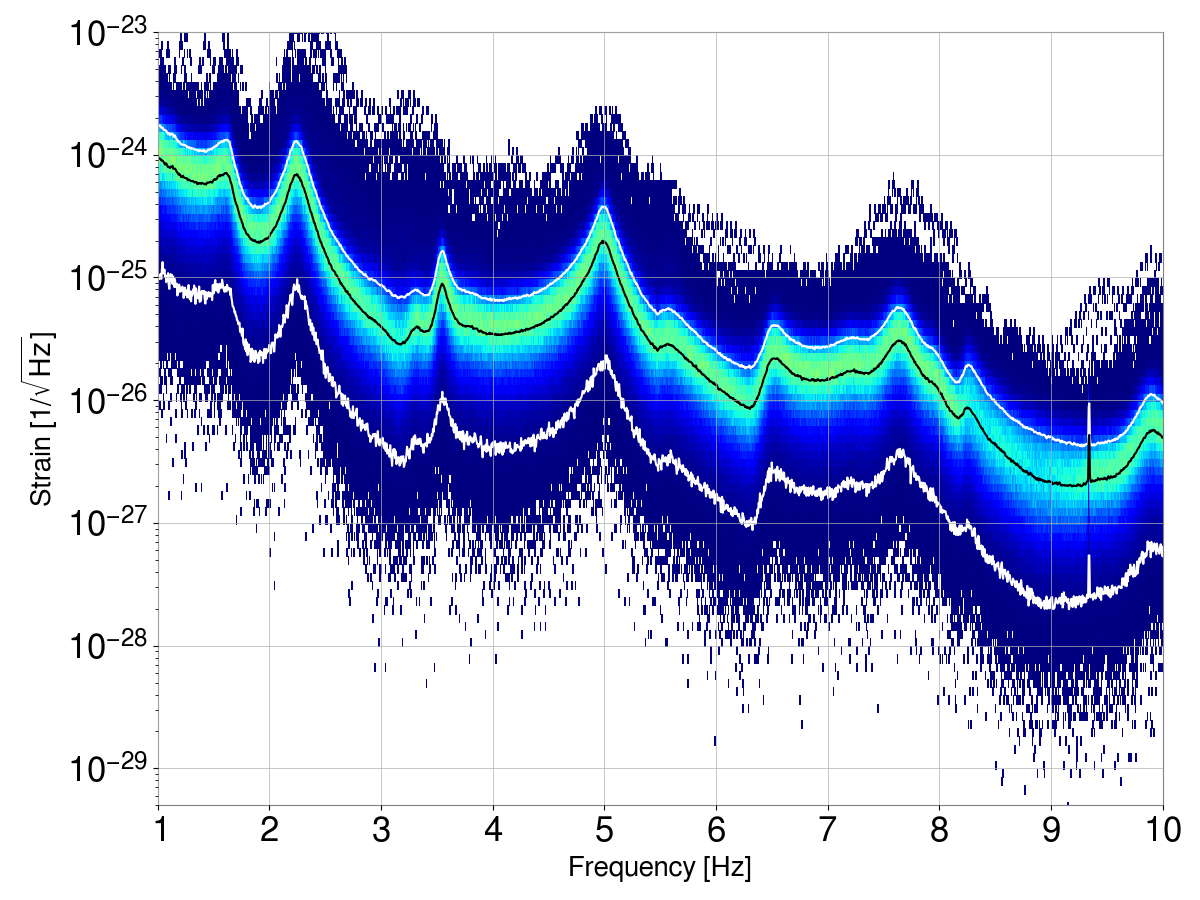}
	 \end{minipage}
	\caption{NN estimate from surface Rayleigh waves - data from GIFH10 sensor (\textit{left}), from body waves - data from position 5 (\textit{right}) and from cavern acoustic fields - data from microphone ACO TYPE 7146NL (\textit{bottom}). The black lines are the 50$^{th}$ percentile, while the white lines represent the 10$^{th}$ and the 90$^{th}$ percentiles.}
\label{fig:KAGRA-NN-estimated}
\end{figure}

For completeness, we also calculate the Rayleigh-wave NN contribution using data from the GIFH10 seismic data and the equation \cite{Harms-review}:
\begin{equation}\label{eq:NN-Rayleigh}
S^R\left(\delta a_x; \omega\right)= (2\pi G\rho_0\gamma(\nu)e^{-h\omega/v})^2\frac{1}{2}S\left(\xi^R; \omega\right),
\end{equation}
where $G$ is the gravitational constant, $\gamma=0.8$ accounts for the suppression of NN due to sub-surface (de)compression of soil by Rayleigh waves. This parameter depends on ground properties. For the average density of the medium we use $\rho=3000\,\rm kg/m^3$. $S(\xi;\omega)$ is the PSD of vertical surface displacement, and $h=200$\,m is the distance of the test mass to the surface. Note that KAGRA's end stations are about 450\,m deep, which means that Rayleigh NN will contribute more at the corner station. Finally, for the velocity we use $v = 3.2$\,km/s corresponding to the median value at 1\,Hz in \autoref{fig:KAGRA-velocities}. This means that we likely over-estimate Rayleigh NN since the waves in the NN band 5\,Hz -- 20\,Hz will be significant slower than at 1\,Hz, which would mean higher NN suppression with depth. 
	
\begin{figure}[h!]
	\begin{minipage}{0.9\textwidth}
		\centering
		\includegraphics[width=1\textwidth]{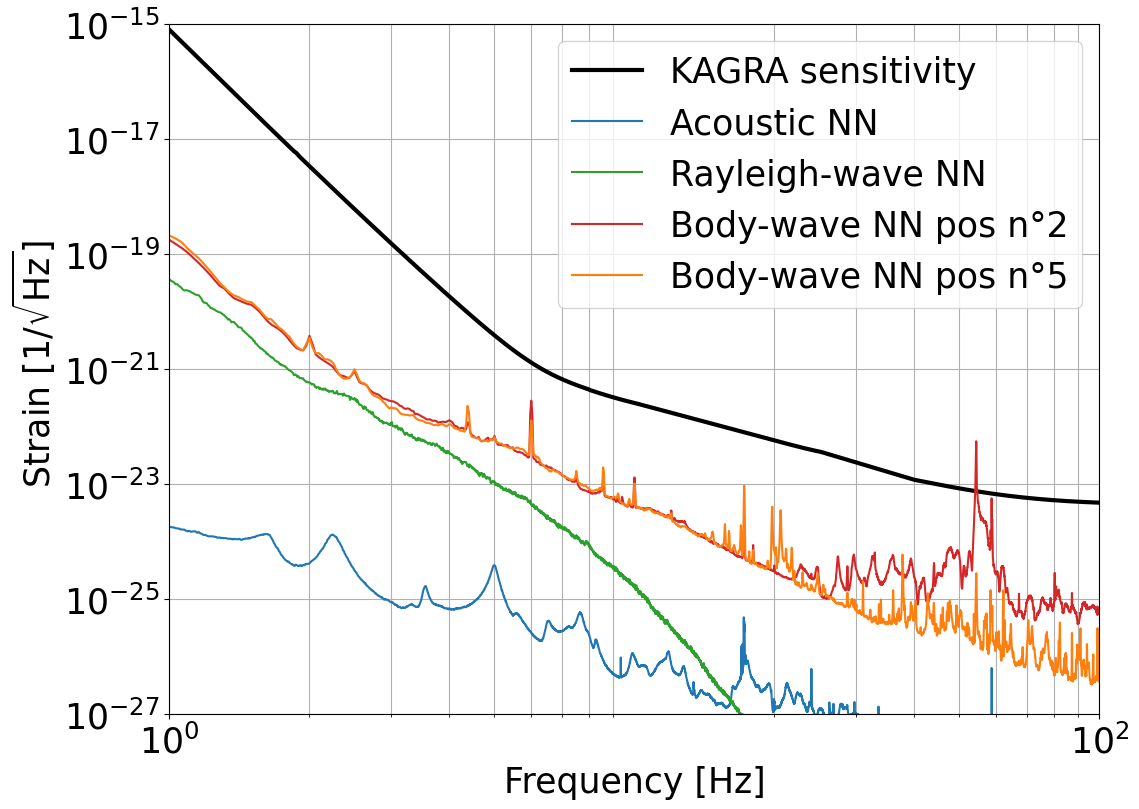}
\caption{Modelled acoustic and seismic NN spectra (90$^{\rm th}$ percentiles). The KAGRA sensitivity curve represents an optimistic scenario (130\,Mpc horizon for neutron-star binaries) for the upcoming science run O4 \cite{Abbott2020-obs}. The seismic spectrum used for the pos 5 NN spectrum is not representative of conditions during observing runs. It is only shown to illustrate what effect infrastructure might potentially have on NN.}
	\label{fig:KAGRA-NN-comparison1}
	\end{minipage}
\end{figure}

\begin{figure}
    \centering
    \includegraphics[width=0.7\textwidth]{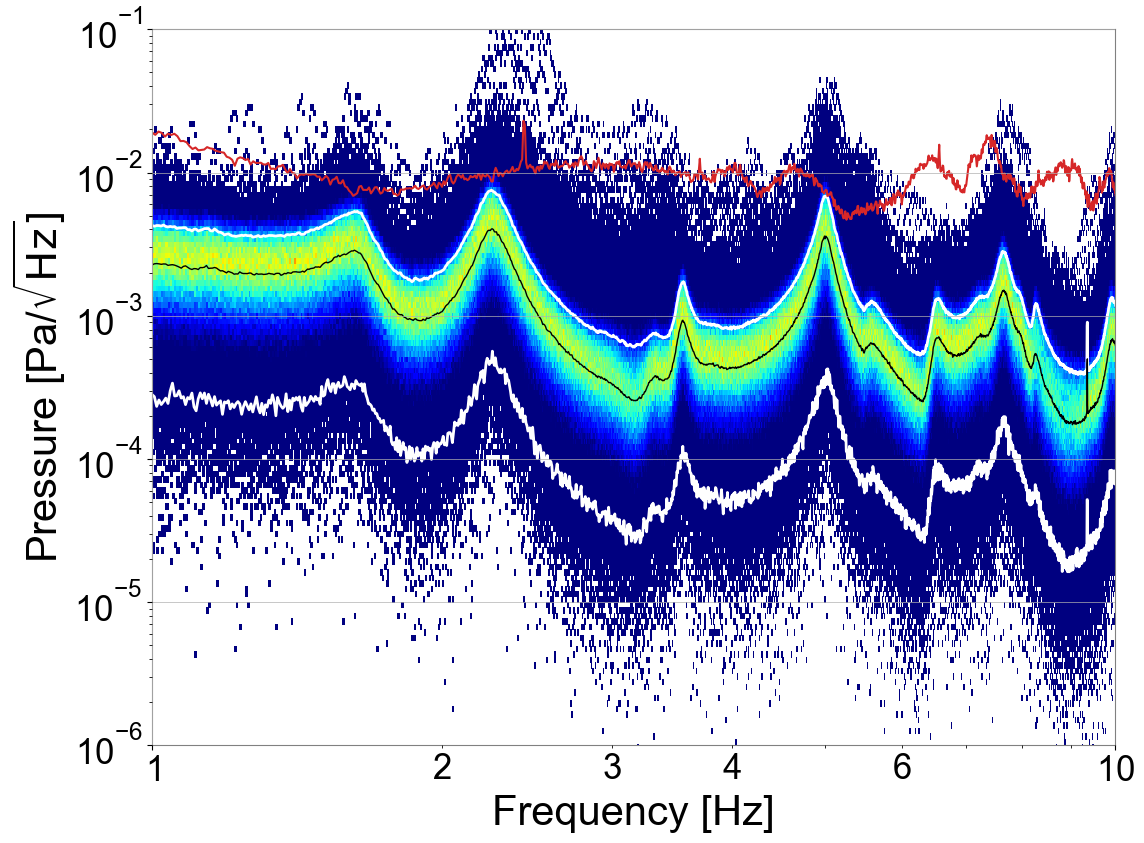}
    \caption{Acoustic noise in Virgo (red line) compared to the acoustic noise recorded by the microphone (ACO TYPE 7146NL) close to the BS (see \autoref{fig:seismometers-positions}). The black line is the 50$^{th}$ percentile, the white lines represent the 10$^{th}$ and the 90$^{th}$ percentiles.}
    \label{fig:mic}
\end{figure}

In \autoref{fig:KAGRA-NN-estimated} the NN histograms coming from Rayleigh waves (left), body waves (right), and the acoustic field in the caverns (bottom) are shown. The acoustic NN is based on a measurement close to the BS (see \autoref{fig:mic}). 

The model employed for the acoustic NN is equation (13) of \cite{AmEA2020} with the assumption that the acoustic NN budget in the end stations was the same as in the corner station:
\begin{equation}
S^h_{\rm cav}(f) = \left( \frac{2c_sG\rho_0\delta p_{\rm cav}(f)}{p_0\gamma f}\right)^2\frac{1}{3}(1-\text{sinc}(2\pi f R/c_s))^2\frac{4}{L^2(2\pi f)^4}     
\end{equation}
Where $c_s = 340$ m/s is the speed of the sound, $\gamma = 1.4$ the adiabatic coefficient and $\rho_0$ and $p_0$ the mean air density and pressure. $L$ is the length of KAGRA's arms (3 km) and $R$ is the cavern radius assumed to be 10\,m, which represents the largest radius of the non-spherical cavern, and which therefore leads to a conservative acoustic NN estimate. The acoustic NN spectrum has more structure, possibly because it is produced by sources located inside the cavern and because resonances can form to enhance the amplitude of acoustic waves at certain frequencies. \autoref{fig:KAGRA-NN-comparison1} shows a comparison of the spectra. We can see that Rayleigh-wave NN is always lower than body-wave NN, even if close to 3\,Hz they start to be similar, while the acoustic NN generated by the infrastructure is significantly weaker than body-wave NN throughout the frequency band of interest.

In order to have a better idea of a possible impact of infrastructure noise on NN spectra, we also show a body-wave NN estimate in \autoref{fig:KAGRA-NN-comparison1} based on the seismic spectrum measured at position 2. As mentioned before, the excess noise that we observed at position 2 is not present during KAGRA's science runs, but we use it to investigate whether such a peak could pose a sensitivity limitation in the form of NN. Indeed, the peak in the spectrum reaches above the KAGRA sensitivity model between 50\,Hz and 60\,Hz. It is clear that such strong infrastructure noise must be avoided.

\section*{Conclusions}
\label{sec:concl}
In this article,  we presented an analysis of seismic spectra observed at the KAGRA underground site, and we used these spectra to estimate Newtonian noise (NN) in the KAGRA detector. Three seismometers located at the two end stations and the corner station of the KAGRA detector were used to estimate speeds of Rayleigh waves at the site between about 0.3\,Hz and 1\,Hz. The estimate was used for a (very approximate) prediction of Rayleigh-wave NN whose amplitude is attenuated with depth depending on the length of Rayleigh waves. 

A comparison of NN spectra including an estimate of NN from the acoustic field in the caverns showed that seismic body-wave NN is the strongest contribution. However, even body-wave NN lies well below the KAGRA sensitivity target. Only in a narrow frequency band between 50\,Hz and 60\,Hz, excess seismic NN produced by the infrastructure might lead to a sensitivity limitation. 

The analyses of seismic spectra provide crucial insight for the Einstein Telescope infrastructure design. First, seismic spectra seem to be unperturbed by infrastructure sources below 20\,Hz. Second, the seismic excess noise observed above 20\,Hz at one of the measurement locations quickly attenuated. Already at a distance of 35\,m (the distance between position 2 and position 3), excess noise was not visible in the data anymore. These findings lead to two preliminary conclusions, which need to be tested further: (1) Infrastructural sources at KAGRA do not deposit significant energy into the seismic field. This is in stark contrast to the surface sites of the Virgo and LIGO detectors, where infrastructural noise is known to dominate the seismic field in the NN band. Whether the machines at KAGRA (pumps, compressor, ventilation, ...) are relatively silent, or whether the hard rock helps to reduce the vibration coupling to ground needs to be investigated with additional seismic measurements. (2) The quick attenuation of seismic excess noise with distance indicates that it originates from the acoustic field in the caverns. It is possible that pressure fluctuations in the caverns act on the ground (or directly on the seismometers) to produce this noise. These ground vibrations are not described by seismic waves (since their amplitude would not attenuate so quickly), but by local response to cavern pressure fluctuations. If this is a generic feature of underground sites in hard-rock environments, then it would mean that infrastructural excess noise can be easily avoided in the vicinity of test masses. Follow-up studies are necessary at the KAGRA site to corroborate these preliminary findings.


\section*{Acknowledgements}
This research has made use of data, software and web tools obtained or developed by the KAGRA Collaboration. The KAGRA project is funded by the Ministry of Education,Culture,Sports,Science and Technology (MEXT) and the Japan Society for the Promotion of Science (JSPS). We acknowledge support from the JSPS Core-to-Core Program A, and the Advanced Research Networks, JSPS Grant-in-Aid for JSPS Fellows 19J01299 and Joint Research Program of the Institute for Cosmic Ray Research (ICRR) University of Tokyo 2019-F14, 2020-G12, and 2020-G21. The work of FB, IF, and FP was supported by funding from the European Union’s Horizon 2020 research and innovation programme under the Marie Skłodowska-Curie grant agreement No 734303 and No 101003460. We acknowledge the use of seismic data from the Hi-net seismometer network (station GIFH10).

\newpage
\section*{References}
\bibliography{myref}
\bibliographystyle{iopart-num}
\end{document}